\documentclass[twocolumn,epjc3]{svjour3} 
\RequirePackage[T1]{fontenc}
\smartqed
\RequirePackage{graphicx}
\RequirePackage{mathptmx}
\RequirePackage{flushend}
\RequirePackage[numbers,sort&compress]{natbib}
\RequirePackage[colorlinks,citecolor=blue,urlcolor=blue,linkcolor=blue]{hyperref}
\usepackage{amsmath}
\journalname{Eur. Phys. J. C}

\begin{document}

\title{Exact Solutions and Accelerating Universe in Modified Brans-Dicke Theories}
\author{Purba Mukherjee\thanksref{e1,addr1}
        \and
        Soumya Chakrabarti\thanksref{e2,addr2} 
}
\thankstext{e1}{e-mail: pm14ip011@iiserkol.ac.in}
\thankstext{e2}{e-mail: soumya@cts.iitkgp.ernet.in}
\institute{{Department of Physical Sciences,~~\\Indian Institute of Science Education and Research Kolkata,\\ Mohanpur, West Bengal 741246, India.}\label{addr1}\\
          \and 
          {Centre for Theoretical Studies,~~\\Indian Institute of Technology Kharagpur,~~\\Kharagpur 721 302, West Bengal, India}\label{addr2}
}
\date{Received: date / Accepted: date}
\maketitle

\begin{abstract}
Exact solutions are studied in the context of modified Brans-Dicke theory. The non-linearity of the modified Brans-Dicke field equations is treated with the Euler-Duarte-Moreira method of integrability of anharmonic oscillator equation. While some solutions show a forever accelerating nature, in some cases there is a signature flip in the evolution of deceleration parameter in recent past. Importance of these latter models are studied in the context of late time acceleration of the universe. Constraints on the model parameters are obtained from Markov Chain Monte Carlo (MCMC) analysis using the Supernova distance modulus data, observational measurements of Hubble parameter, Baryon acoustic oscillation data and the CMB Shift parameter data.
\end{abstract}


\vspace{1cm}
\newcommand{\be}{\begin{equation}}
\newcommand{\ee}{\end{equation}}
\newcommand{\bea}{\begin{eqnarray}}
\newcommand{\eea}{\end{eqnarray}}

\section{Introduction}
Introduced back in $1915$ \cite{einstein}, General Theory of Relativity (GR) has remained the most successful description of gravity till date. Based on an equivalence between gravitation and inertia, GR ensures that one can write down the geodesics on a spacetime from the metric structure \cite{will}. The foundation principles of the theory has been well tested over the years through famous experiments, for instance, the E\"{o}tv\"{o}s experiment \cite{eotvos}, Michelson-Morley-type experiments \cite{mm, brill}, and the gravitational redshift experiments \cite{will1}. Moreover, recent observational evidences tend to confirm the theoretically predicted outcomes of GR as well, for example, the existence of black holes and gravity waves \cite{ligo}, promotes GR as a potentially rich store of possibilities even in the current context. \\

In spite of passing many theoretical and experimental tests successfully, there remains some unresolved issues hinting towards a possible modification of GR. An immediate motivation comes from the recently discovered late time accelerated expansion of the universe. To account for such a non-trivial acceleration, the simplest possible modification is to consider a cosmological constant $\Lambda$ playing the role of `dark energy', a fluid responsible for an effective negative pressure. However, this option is problematic. Defining an empty space as a collection of quantum fields and assuming that the zero-point fluctuations of such quantum fields contributes to $\Lambda$, the theoretically predicted vacuum energy scale overwhelmingly mismatches with the observed vacuum energy \cite{weinberg}. A very well-studied alternative is to consider a time-dependent scalar field acting as the generator of the non-trivial acceleration, for example, the quintessence models or the phantom scalar fields (for extensive details on scalar field cosmology, see for instance, Ratra and Peebles \cite{ratra}, Brookfield et. al. \cite{brook}, Overduin and Cooperstock \cite{over}, Bento, Bertolami and Sen \cite{sen} and references therein). Another possibility is to treat the dark energy component as a geometric quantity, coming out of a modified Einstein-Hilbert action. For example, replacing the Ricci curvature $R$ in the Einstein-Hilbert action by a general function $f (R)$ produces the so-called $f (R)$ theories. For extensive reviews on different modifications of gravity, we refer to the works of Nojiri and Odintsov \cite{nojiodi}, Capozziello and De Laurentis \cite{delaur}, Faraoni and Capozziello \cite{faracapo}, Bamba and Odintsov \cite{bamba1}, Nojiri, Odintsov and Oikonomou \cite{nojiodioik}.  \\

Brans-Dicke (BD) theory is a modification of gravity which was formulated in order to incorporate the Mach's Principle in GR \cite{bd}. Mach's principle simply demands that the local motion of a particle must be affected by a large-scale matter distribution. This can be violated in GR, for example in the de-Sitter universe dominated by a cosmological constant where matter is entirely negligible \cite{misner}. In BD theory, a scalar field $\phi$ is included in the action which makes the gravitational coupling a function of coordinates, rather than a constant. A dimensionless parameter $\omega$, called the BD parameter, governs the departure of the results obtained under weak field approximation of the theory from that of general relativity. To be consistent with the local astronomical tests, $\omega$ must have a very high value ($\omega > 500$). It can be proved that in the limit $\omega \rightarrow \infty$, $\phi$ reaches a constant value $\sim \frac{1}{\omega}$, making the field equations of BD theory equivalent to the corresponding GR equations \cite{weinbook}. Despite carrying this great advantage of giving back GR in some limit, it was eventually proved that the infinite $\omega$ limit fails for a traceless stress-energy distribution, by Banerjee and Sen \cite{banerjeesen}, Faraoni \cite{faraoni}. \\

However, BD theory remains well regarded as the prototype of a large class of theories of gravity called the scalar-tensor theories, where a non-minimal coupling exists between a scalar field and the curvature scalar. These theories receive significant attention cosmologically as they can successfully describe an early inflation of the universe \cite{guth}. BD theory predicts a period of extended inflation to tackle the graceful exit problem of the universe as described by Mathiazhagan and Johri \cite{mathi}, La and Steinhardt \cite{la}. Moreover, the theory can successfully generate the late time accelerated expansion as well, for suitable values of the parameter $\omega$ without the need of any exotic matter field (see for instance, Banerjee and Pavon \cite{nbpavon1}). Introduction of an additional self-interaction potential of the BD scalar field forms a straightforward modification of the theory and is well-studied in literature, for example by Faraoni and Gunzig \cite{faragun}, Bertolami and Martins \cite{bertolami}. Banerjee and Pavon proved that the theory can produce a non-decelerated expansion in the presence of an additional minimally coupled scalar field \cite{nbpavon2}. Sen and Sen looked into the possibility of a late time acceleration in BD theory for some specific choice of an additional potential term \cite{sensen}. Das and Banerjee showed the possibility of a late time accelerated expansion preceded by a decelerated expansion in the domain of the theory, considering a non-minimal coupling of matter sector and the BD scalar field \cite{dasbanerjee}. Cosmological solutions in BD theory were recently discussed by Jarv, Kuusk, Saal and Vilson for both Einstein and Jordan frames \cite{jarv}. Generalizing the theory by making $\omega$ a function of the scalar field is another interesting possibility, proposed by Bergmann \cite{berg}, Wagoner \cite{wago} and Nordtvedt \cite{nord}. Indeed, non-minimally coupled scalar-tensor theories (for example, the works of Barker \cite{barker} and Schwinger \cite{schwinger}) become equivalent to a generalized BD theory for a suitable choice of $\omega(\phi)$. For a brief review on such modifications and their cosmological motivations we refer to the work of Van den Bergh \cite{vdb}. For recent discussions on BD theory and more generally, scalar tensor theories and their cosmological implications we refer to the works of Fujii and Maeda \cite{maeda}, Sotiriou \cite{soti}.\\

In the present work, we focus on exact solutions in modified BD theory. The specific modified BD actions we study has previously been studied in literature (\cite{clifbarrow, dasbanerjee, nbpavon2}). However, exact solutions have not really been considered. Exact solutions are never guaranteed mainly due to the high degree of non-linearity of the BD field equations, however, they remain an interesting avenue of research carrying a pedagogical importance. Taking a different approach from assuming a cosmic expansion history at the outset, we incorporate a purely mathematical property of general second order differential equations with variable coefficients that can be point transformed into an integrable form. The property is derived from the symmetry analysis of classical anharmonic oscillator equations by Duarte, Moreira, Euler and Steeb \cite{duarte}; generalized by Euler, Steeb and Cyrus \cite{euler1} and thereafter expanded by Euler \cite{euler2}, Harko, Lobo and Mak \cite{harko}. The exact solutions extracted using this property give accelerating cosmological solutions. While some of the solutions give forever accelerating solutions, some examples indeed show a signature flip of deceleration parameter in recent past, hinting that such models can work well to model a late time acceleration. For relevant models we perform a Markov Chain Monte Carlo (MCMC) analysis using the supernova distance modulus data, observational measurements of Hubble parameter, baryon acoustic oscillation data and the CMB Shift parameter data to study the constraints on the model parameters. We reconstruct the cosmological quantities for the best fit values of the model parameters. \\

We consider two different modifications of the standard BD action. The first modifiction involves a non-minimal coupling of the matter sector with the BD scalar field. A similar action finds it's existence in the low energy limit of string theory or the dilaton gravity. Under such a setup, the BD scalar field behaves as a chameleon scalar field (for an idea and some examples of chameleon fields and their relevance in cosmology, we refer to the works of Khoury and Weltman \cite{khoury1, khoury2}, Mota and Barrow \cite{mota1, mota2}, Das, Corasaniti and Khoury \cite{dck}, Mota and Shaw \cite{motashaw1, motashaw2}). Clifton and Barrow \cite{clifbarrow} studied the cosmological solutions and their relevance for a similar setup of BD theory. Das and Banerjee \cite{dasbanerjee} studied accelerating solutions in this setup assuming cosmic expansion history at the outset and discussed the evolution of deceleration parameter for different epochs. In the present work, instead of using any apriori ansatz over the scale factor or the BD scalar field, we solve the modified BD field equations for a general power-law non-minimal coupling between matter and scalar field. The second modification involves a minimally coupled self-interacting scalar field serving as a quintessence matter within the standard action of BD theory. A quintessence matter is added to enlarge the scope of the theory since this matter field in known to describe accelerated expansion under the scope of standard GR itself, confirmed by the observational data from the luminosity-redshift relation of type $I$a supernovae \cite{perl}. Amongst many scalar field models considered as quintessence matter in literature, the model proposed by Zlatev, Wang and Steinhardt \cite{zws} serves particular importance, where a tracker field slowly rolls down the potential. The role of such a scalar field in BD theory was studied by Banerjee and Pavon \cite{nbpavon2} for some particular choices of the self-interaction potential. We study exact solutions for different self-interaction potentials, a simple power law and the other being a combination of power law terms. For a self-interaction potential of the form $V(\phi) \sim \phi^{\delta_1} + \phi^{\delta_2}$, the integrability property of the scalar field equation produces some interesting exact solutions hinting at a late-time accelerated expansion. \\

In section $2$, we briefly outline the mathematics of point transformation and direct integration of a general anharmonic oscillator equation. In section $3$ we present exact solutions describing a forever accelerationg cosmology for some modifications of standard BD theory. Section $4$ contains exact solutions describing a late time acceleration of the universe for a Quintessence plus BD scalar field setup. Solutions present in Section $4$ also closely resemble the observational data as checked in Section $5$ through parameter estimation by statistical analysis and the analysis of the evolution of cosmological parameters. We conclude the manuscript in section $6$.	

\section{Integrability of Anharmonic Oscillator Equations}
An anharmonic oscillator can be written as a second order differential equation having variable coefficients.
\begin{equation}\label{anharmonic}
\ddot{\phi}+f_1(t)\dot{\phi}+ f_2(t)\phi+f_3(t)\phi^n=0.
\end{equation}
Here, $f_1$, $f_2$ and $f_3$ are variable coefficients which are functions of $t$ and $n$ is a constant. This equation can be written in an integrable form by a pair of point transformations, if and only if $n\notin \left\{-3,-1,0,1\right\} $. Additionally, the coefficients must satisfy the differential condition

$$ \frac{1}{(n+3)}\frac{1}{f_{3}(t)}\frac{d^{2}f_{3}}{dt^{2}} - \frac{(n+4)}{\left( n+3\right) ^{2}}\left[ \frac{1}{f_{3}(t)}\frac{df_{3}}{dt}\right] ^{2} $$
\be \label{criterion1}
~~~~~~~~~~~~~~~~~~~~~+ \frac{(n-1)}{\left( n+3\right) ^{2}}\left[ \frac{1}{f_{3}(t)}\frac{df_{3}}{dt}\right] f_{1}\left( t\right) + 
\ee
$$ \frac{2}{(n+3)}\frac{df_{1}}{dt} +\frac{2\left( n+1\right) }{\left( n+3\right) ^{2}}f_{1}^{2}\left( t\right)=f_{2}(t). $$

The point transformations are to be defined as 
\begin{eqnarray}\label{critertion2}
\Phi\left( T\right) &=&C\phi\left( t\right) f_{3}^{\frac{1}{n+3}}\left( t\right)
e^{\frac{2}{n+3}\int^{t}f_{1}\left( x \right) dx },\\
T\left( \phi,t\right) &=&C^{\frac{1-n}{2}}\int^{t}f_{3}^{\frac{2}{n+3}}\left(
\xi \right) e^{\left( \frac{1-n}{n+3}\right) \int^{\xi }f_{1}\left( x
\right) dx }d\xi ,\nonumber\\
\end{eqnarray}%
where $C$ is a constant. \\

The scope of this approach in gravitational physics has been studied only very recently, for example, in massive scalar field collapse \cite{scnb1, scnb2} and Scalar-Einstein-Gauss-Bonnet gravity \cite{sc}, cosmological reconstruction of modified theories of gravity \cite{scjs, sckbjs}.

\section{Exact Solutions and forever accelerating cosmologies}

\subsection{{\bf Brans Dicke Scalar Field as a Chameleon (Model $I$)}}
In this section we work with a setup where the BD scalar field $\phi$ is non-minimally coupled to the matter distribution through an arbitrary function $f(\phi)$. The relevant action is written as

\be \label{action1}
\mathit{A}=\frac{1}{16 \pi } \int \sqrt{-g} \left[ \phi R -\frac{\omega}{\phi} {\phi_{,\mu}}{\phi^{,\mu}} + \mathit{L_m}f\left( \phi \right) \right] d^4x,
\ee

where $\omega$ is the Brans Dicke parameter. $R$ is the standard Ricci scalar and $\mathit{L_m}$ defines the matter distribution which we assume to be pressureless dust. One may note that for $f\left( \phi \right) = 1$, one gets back the usual BD action. We define the metric for a homogeneous and isotropic, spatially flat universe as

\be \label{metric}
ds^2 = dt^2 - a^2(t) \left[ dr^2 +r^2d{\Omega}^2 \right],
\ee
where $a(t)$ is the scale factor of the universe. Variation of the action with respect to the metric gives the field equations as,
\be
3 {\left(\frac{\dot{a}}{a}\right)}^2 =  \frac{\rho_m f}{\phi}  - 3 \frac{\dot{a}}{a} \frac{\dot{\phi}}{\phi}+ \frac{\omega}{2} {\left( \frac{\dot{\phi}}{\phi} \right)}^2 ,
\label{EE0a}
\ee

\be
2\frac{\ddot{a}}{a} + { \left( \frac{\dot{a}}{a} \right)}^2 = -\frac{\omega}{2} {\left(\frac{\dot{\phi}}{\phi} \right)}^2 - 2 \frac{\dot{a}}{a} \frac{\dot{\phi}}{\phi} - \frac{\ddot{\phi}}{\phi} . \label{EE1a}
\ee

A \textit{dot} denotes derivative with respect to cosmic time $t$. $\rho_{m}$ denotes density of the fluid distribution. The condition for conservation of energy-momentum distribution for a dust fluid can be written as
\be \label{matter1a}
\dot{\rho_m} + 3 \frac{\dot{a}}{a} \rho_m = -\frac{3}{2}\frac{\dot{f}}{f}.
\ee
Varying the action with respect to the BD scalar field $\phi$ one can also write

\be \label{phi1a}
\ddot{\phi} + 3 \frac{\dot{a}}{a} \dot{\phi} = \frac{\rho_m}{\left( 2\omega +3 \right)} \left[ f+\frac{df}{d \phi} \phi \right]. 
\ee

From equation (\ref{matter1a}) one arrives at the evolution of matter density as a function of the scale factor, written as
\be \label{matter2a}
\rho_m = \frac{\rho_0}{a^3 f^{\frac{3}{2}}},
\ee
where, $\rho_0$ is a constant of integration. One can note from equation (\ref{matter2a}) that due to the presence of the non-minimal coupling function $f(\phi)$, there is a departure from usual matter conservation equation. Equation (\ref{matter2a}) and (\ref{phi1a}) together produces the following differential equation as

\be \label{phifinala}
\ddot{\phi} + 3 \frac{\dot{a}}{a} \dot{\phi} = \frac{\rho_0}{{\left( 2\omega +3 \right)}a^3} \left[ f^{-\frac{1}{2}}+\frac{df}{d \phi} f^{-\frac{3}{2}} \phi \right].
\ee

We study the equation (\ref{phifinala}) for a power law coupling function, i.e., $f(\phi) \sim \phi^\delta$. Under such an ansatz for $f(\phi)$ equation (\ref{phifinala}) falls within the large class of equations that can be identified as a classical anharmonic oscillator equation. These equations can be point-transformed into an integrable form. We now consider a power law coupling such that $f(\phi)$ can be written as $f(\phi)= {\phi}^{m}$. With this the equation (\ref{phifinala}) can be written as
\be
\ddot{\phi} + 3\frac{\dot{a}}{a} \dot{\phi} - \frac{\rho_0}{a^3} \left( \frac{1+m}{2 \omega + 3} \right) {\phi}^{-\frac{m}{2}} = 0.
\ee

We write $n$ as $-\frac{m}{2}$. Therefore, 
\be
\ddot{\phi} + 3\frac{\dot{a}}{a} \dot{\phi} + \frac{\rho_0}{a^3} \left( \frac{2n-1}{2 \omega + 3} \right) {\phi}^{n} = 0.
\ee

One may note that this equation indeed is a particular case of anharmonic oscillator with the coefficients identified as,
\begin{eqnarray}
f_1 & = & 3 \frac{\dot{a}}{a}, \\
f_2 & = & 0 ,\\
f_3 & = & \rho_0 \left( \frac{2n-1}{2\omega +3} \right) \frac{1}{a^3} = \frac{D}{a^3}.
\end{eqnarray}

Therefore one can transform the above equation into an integrable form and integrate for the BD scalar field. Moreover, the integrability criterion (\ref{criterion1}) is expected to reduce into a solvable equation of the cosmological scale factor. For the analysis to stand, $n$ can not be $-3, -1, 0 ,1$. This implies some restrictions over the choice of $m$ as $m \neq 6, 2, 0, -2$ i.e., $f(\phi) \neq {\phi}^6, {\phi}^2, 1, {\phi}^{-2}$. From equation (\ref{criterion1}), we find an exact solution for the scale factor which goes as,
\be \label{a}
a(t) = {\left\lbrace \frac{3(n+2)}{n+3} \lambda (t - t_0) \right\rbrace }^{\frac{(n+3)}{3(n+2)}}
\ee

From the expression of the scale factor (\ref{a}) it is evident that the nature of the exponent governs the evolution. Now using the condition (\ref{critertion2}) we point transform the scalar field evolution equation (\ref{phifinala}) and solve for the BD field. The solution can also be written from (\ref{phifinala}) simply by using the solution for scale factor as in (\ref{a}) and is given by,
\be \label{phi_Mod1}
\phi(t) = \phi_0 {\left\lbrace \frac{\rho_0(2n-1)}{2 \omega +3} \right\rbrace }^{\frac{1}{1-n}} {\left( t -t_0 \right)}^{-\frac{(n+1)}{(n+2)(n-1)}}.
\ee
where, $\phi_0$ is to be determined from the consistency check of the theorem. Amongst the equations (\ref{EE0a}), (\ref{EE1a}), (\ref{matter1a}) and (\ref{phi1a}), only three are independent equations as the fourth can always be derived using the Bianchi identity. Therefore an exact solution coming out from (\ref{matter1a}) and (\ref{phi1a}) is a consistent solution as long as it satisfies any one of the field equations. From Eq. (\ref{EE1a}) the consistency criterion can be written as  

\begin{eqnarray}\nonumber \label{consistency}
&&\omega = -\frac{2(n+2)^{2}(1-n)^{2}}{(n+1)^{2}} \Bigg[\frac{(n+3)^{2}}{3(n+2)^{2}} - \frac{2(n+3)}{3(n+2)} \\&&\nonumber
+ \frac{2(n+3)(n+1)}{3(n+2)^{2}(1-n)} + \frac{(n+1)^{2}}{(n+2)^{2}(1-n)} - \frac{(n+1)}{(n+2)(1-n)}\Bigg].
\end{eqnarray}

\begin{figure} [h!]
\begin{minipage}{\columnwidth}
\centering
		\includegraphics[angle=0, width=0.8\textwidth]{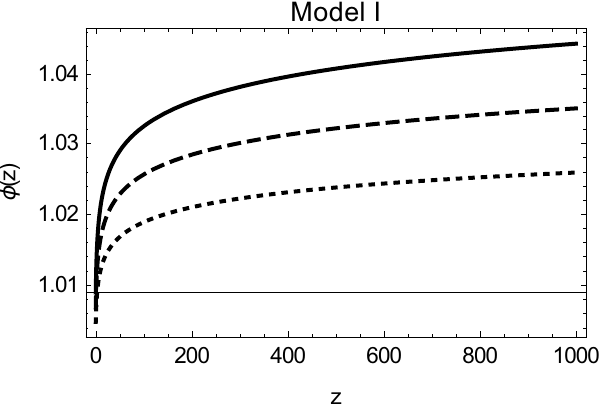}
\end{minipage}
\begin{minipage}{\columnwidth}
\centering
		\includegraphics[angle=0, width=0.78\textwidth]{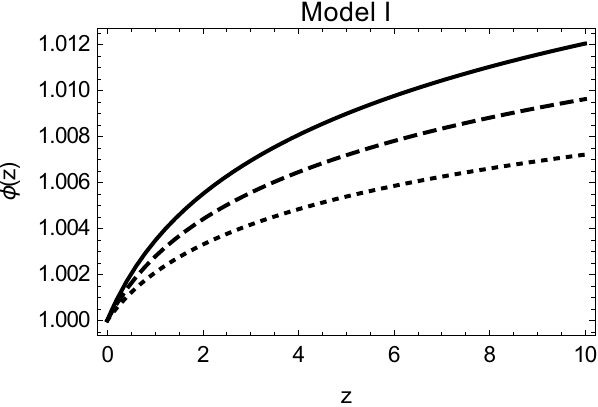}
\end{minipage}
\begin{minipage}{\columnwidth}
\centering
		\includegraphics[angle=0, width=0.82\textwidth]{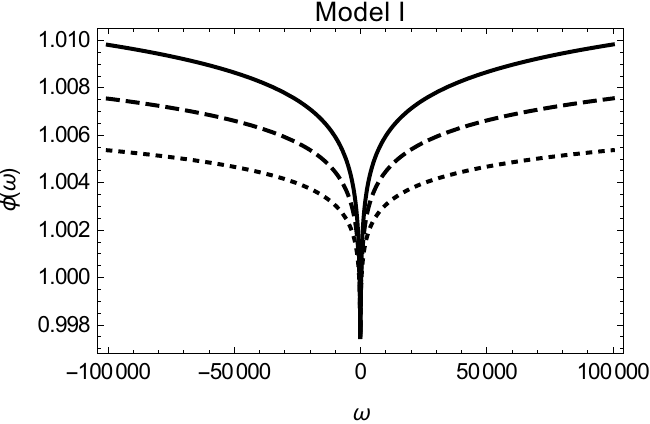}
\end{minipage}
\caption{{\small Plots of the Brans Dicke scalar field $\phi$ for Model $I$ using different $n$. The top most figure shows the logarithmic variation of $\phi$ as a function of $z$ scaled with $\phi_0$. The graph in the middle shows the scaled out variation of $\phi$ w.r.t. $z$ at low redshifts. The lowermost graph gives the variation of $\phi$ as a function of the Brans Dicke parameter, $\omega$. In the figure, \textit{$n = 650$ : \textbf{Black}, $n = 750$ :\textbf{ Black, Dashed}, $n = 1000$ : \textbf{Black, Dotted}}.}}
	\label{phiM1plot}
\end{figure}

For a numerical analysis, the value of $\rho_0$ is obtained from the relation of matter density parameter, $\Omega_{m0}$. We define $\Omega_{m0} = \frac{8 \pi G \rho_0}{3{H_0}^2}$, where we use the value of $\Omega_{m0} \approx 0.3$ \cite{planck} and $H_0$ = $72.8$ $ \mbox{km Mpc}^{-1} \mbox{sec}^{-1}$ \cite{riess2018}, being consistent with recent observational data. The value of $\omega$ considered is nearly $60,000$ for the results to be consistent with the local astronomical tests. We choose a suitable value of $n$ following Eq. (\ref{consistency}) such that this domain of $\omega$ is strictly maintained. Fig. \ref{phiM1plot} shows the logarithmic variation of $\phi$ (top graph) as a function of $z$ scaled with $\phi_0$. The graph in the middle gives the variation of $\phi$ w.r.t. $z$ at low redshifts. The lowermost graph gives the variation of $\phi$ as a function of the Brans Dicke parameter $\omega$.  \\

We note that the deceleration parameter for Model $I$ is a constant for all $z$, given by $\left(\frac{2n+3}{n+3}\right)$. Therefore, depending on the respective $n$ value we will get either a constant accelerating or decelerating universe. Thus, Model $I$ becomes redundant as far as present observational cosmology is concerned. However, it still serves as a simple toy model based on simple power-law potential assumption, even more so because the scale factors they produce are very simple. Moreover, a power law evolution of scale factor $a(t) \sim t^{\alpha}$ may have more interesting role in early universe cosmology. 

\subsection{{\bf Brans Dicke with a Quintessence and Simple Power Law Potential (Model  $II$)}}
In this section we study the standard BD action alongwith a self-interacting scalar field playing the role of Quintessence matter. The corresponding action is given by,

\be \label{action2}
\mathit{S}=\frac{1}{16 \pi } \int \sqrt{-g} \left[ \phi R -\frac{\omega}{\phi} {\phi_{,\alpha}}{\phi}^{,\alpha} + \mathit{L_m} \right] d^4x.
\ee

$\phi$ defines the BD Scalar Field, $\omega$ is the BD parameter and $R$ is the Ricci scalar. $\mathit{L_m}$ is the matter distribution and in the present case, is defined by a combination of a perfect fluid and a spatially homogeneous scalar field $\psi$ with a self-interaction potential.\\

The field equations, where the metric is given by equation (\ref{metric}) are,
\be
3 {\left(\frac{\dot{a}}{a}\right)}^2 =  \frac{\rho_m + \rho_\psi}{\phi}  - 3 \frac{\dot{a}}{a} \frac{\dot{\phi}}{\phi}+ \frac{\omega}{2} {\left( \frac{\dot{\phi}}{\phi} \right)}^2,
\label{EE0}
\ee

\be
2\frac{\ddot{a}}{a} + { \left( \frac{\dot{a}}{a} \right)}^2 = -\frac{p_m+p_\psi}{\phi} -\frac{\omega}{2} {\left(\frac{\dot{\phi}}{\phi} \right)}^2 - 2 \frac{\dot{a}}{a} \frac{\dot{\phi}}{\phi} - \frac{\ddot{\phi}}{\phi}. \label{EE1}
\ee

$\rho_m$ and $p_m$ denotes the density and the pressure for matter sector. $\rho_{\psi}$ and $p_{\psi}$ denotes the density and pressure for the quintessence field. They are written as a function of $\psi$ as

\begin{eqnarray}
\rho_\psi & = & \frac{{\dot{\psi}}^2}{2} + V \left(\psi \right), \\
p_\psi & = & \frac{{\dot{\psi}}^2}{2} - V \left(\psi \right).
\end{eqnarray}

$V = V (\psi)$ defines the self-interaction potential for the Quintessence scalar field. Varying the action with respect to the scalar field $\psi$ one gets the wave equation corresponding to the scalar field $\psi$ as,

\be
\ddot{\psi} + 3 \frac{\dot{a}}{a} \dot{\psi} + \frac{dV}{d\psi} = 0.
\label{evopsi}
\ee

Moreover, varying the action with respect to the BD scalar field, one gets the wave equation corresponding to $\phi$ as

\be
\ddot{\phi} + 3 \frac{\dot{a}}{a} \dot{\phi} = \frac{1}{\left( 2\omega +3 \right)} \left[ \left( \rho_m -3p_m \right) + \left( \rho_\psi - 3p_\psi \right) \right].
\label{evophi}
\ee

The matter conservation equation for the fluid gives 
\be
\dot{\rho_m} + 3 \frac{\dot{a}}{a}\left( \rho_m + p_m \right) = 0.
\ee

It is not unphysical to assume that in the current era $p_m$ can be put to zero, since the universe is dominantly filled with cold matter. Thus the matter conservation equation gives
\be
\label{rhoq}
\rho_m = \frac{\rho_0}{a^3},
\ee
where $\rho_0$ is a constant of integration. We note here that the evolution equation (\ref{evopsi}) for $\psi$ is a simple case of classical anharmonic oscillator for suitable choices of the self-interaction potentials, as is evident from the form of Eq. (\ref{anharmonic}). Therefore we intend to extract as much information as possible from the above set of equations by virtue of the integrability analysis. Eq. (\ref{evopsi}) will provide us with the exact solution for the scale factor and the Quintessence field $\psi$ for which the equation will be integrable. These solutions will in turn be used to determine $\rho_m$ from Eq. (\ref{rhoq}). Finally, using the solutions for $a(t)$, $\psi(t)$ and $\rho(t)$ in Eq. (\ref{evophi}), we expect to have an idea of the time evolution of the BD scalar field $\phi$. In the following subsections, we study the system of equations for some relevant self-interaction potentials of the quintessence scalar field.\\

The first example taken is for a self-interaction potential power law in $\psi$, given by

\be
V(\psi)= \frac{{\psi}^{n+1}}{n+1}.
\ee

Therefore the evolution equation for the Quintessence Eq. (\ref{evopsi}) becomes
\be
\ddot{\psi} + 3 \frac{\dot{a}}{a} \dot{\psi} + \psi^n = 0.
\ee

Straightaway, this can be identified as a simple case of anharmonic oscillator equation. Following the detailed method as in section $II$, the solutions for $a(t)$ and $\psi(t)$ can be written as

\begin{eqnarray}  \label{psi_mod2}
a(t) & = & {\beta \left( t- t_0 \right)}^{\frac{(n+3)}{3(n+1)}}, \\
\psi(t) & = & \xi {\left( t - t_0 \right)}^{\frac{2}{(1-n)}}.
\end{eqnarray}

provided $n\notin \left\{-3,-1,0,1\right\}$. $\xi$ can be evaluated from a consistency check, written as 
\begin{equation}\label{mod2_xi}
\xi = \Bigg[\frac{2(n+3)}{3(n^2 -1)} - \frac{2(n+1)}{(n-1)^2}\Bigg]^{\frac{1}{(n-1)}}.
\end{equation}

\begin{figure} [h!]
\begin{minipage}{\columnwidth}
\centering
		\includegraphics[angle=0, width=0.8\textwidth]{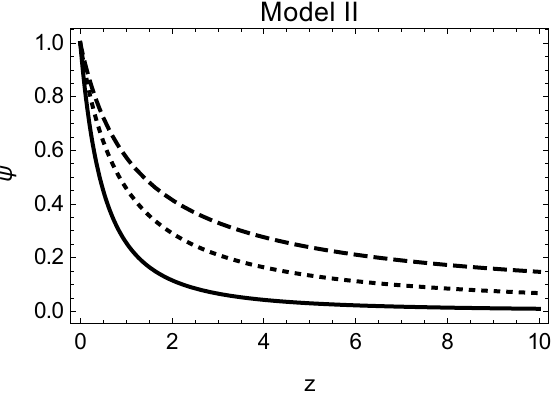}
\end{minipage}
	\caption{{\small Plot of the Quintessence scalar field $\psi$ with redshift for Model $II$ using different values of $n$. In the figure, \textit{$n$ = $-0.01$ : \textbf{Black}, $n$ = $-0.2$ :\textbf{ Black, Dashed}, $n$ = $-0.5$ : \textbf{Black, Dotted}}.}}
	\label{psiM2plot}
\end{figure}

From the expression of the scale factor (\ref{psi_mod2}) we note that for flat cosmology one must have $n > 0$, whereas a negative $n$ very close to zero gives open cosmologies. For $-1 < n < 0$, a late-time acceleration can be realized. Using the exact solutions, and the expression of density (using Eq. (\ref{psi_mod2}) in Eq. (\ref{rhoq})) one gets a second order differential equation for the BD scalar field $\phi(t)$ as,

\begin{figure} [h!]
\begin{minipage}{\columnwidth}
\centering
		\includegraphics[angle=0, width=0.8\textwidth]{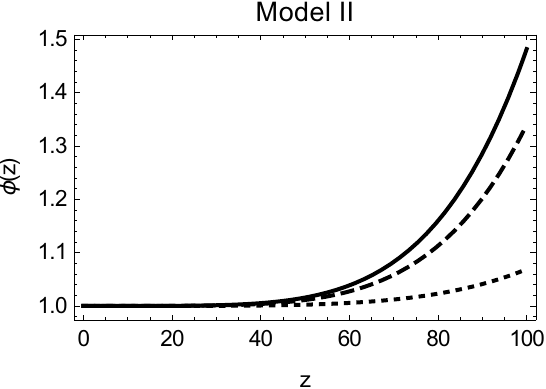}
\end{minipage}
\caption{{\small BD Scalar field for model $II$ as a function of redshift using different values of $n$. In the figure, \textit{$n$ = $-0.01$ : \textbf{Black}, $n$ = $-0.2$ :\textbf{ Black, Dashed}, $n$ = $-0.5$ : \textbf{Black, Dotted}}.}}
	\label{BDfieldmodelz2}
\end{figure}

\begin{eqnarray}\nonumber \label{BDphieq}
&&\ddot{\phi} + \frac{(n+3)}{(n+1)} \frac{\dot{\phi}}{(t-t_0)} - \frac{1}{(2 \omega +3)} \Bigg[ \rho_{0} \Big(\beta(t-t_{0})\Big)^{-\frac{(n+3)}{(n+1)}} \\&&
+ \frac{4 \xi^n}{(1+n)}(t-t_0)^{2 \frac{(1+n)}{(1-n)}} - \frac{4 \xi^2}{(1-n)^2} (t-t_{0})^{2 \frac{(1+n)}{(1-n)}} \Bigg]= 0.
\end{eqnarray} 

Eq. (\ref{BDphieq}) is solved numerically to study the evolution of $\phi$. The evolution as a function of redshift $z$ is shown in Fig \ref{BDfieldmodelz2}, for different values of $n$. The value of $\rho_0$ is obtained from the relation of matter density parameter, $\Omega_{m0}$ just as in case for model $I$. The BD parameter $\omega$ is considered to be $60,000$ to meet the requirement of local astronomical tests. \\

We note that the value of the deceleration parameter for Model $II$ is a constant for all $z$. For Model $II$ it is $\left(\frac{2n}{n+3}\right)$. So, depending on the respective $n$ value we will get either a constant accelerating or decelerating universe. Thus, Model $II$ become redundant as far as present observational cosmology is concerned. However, it can serve as a simple toy model based on simple power-law potential assumptions, even more so because the scale factors they produce are very simple. Moreover, a power law evolution of scale factor $a(t) \sim t^{\alpha}$ may have more interesting role in early universe cosmology. 

\section{Exact Solutions and Late time accelerating cosmologies}

\subsection{{\bf Brans Dicke with a Quintessence and Higgs Potential (Model $III$)}}

The Higgs interaction potential is defined as
\be
V(\psi) = V_0 +\frac{1}{2} {\mu}^2 {\psi}^2 + \frac{1}{4} {\lambda_0} {\psi}^4,
\ee
which is basically a combination of two simple power law self-interaction terms, one a quadratic and another a quartic in $\phi$. This serves an additional purpose as it expands the scope of the integrability analysis. A quadratic potential straightaway does not fall within the scope of the theorem as for a simple $\phi^n$ term in the equation, $n\notin \left\{-3,-1,0,1\right\}$. For the Higgs potential, Eq. (\ref{evopsi}) becomes

\be
\ddot{\psi} + 3 \frac{\dot{a}}{a} \dot{\psi} + \mu^2 \psi + \lambda_0 \psi^3 = 0,
\ee

Analyzing the equation in a similar fashion as in section $II$, we find the exact solution for the scale factor as

\be \label{a_higgs}
a(t) = {\left( \frac{1}{2 {\mu}^2} e^{\sqrt{2}\mu t}-\lambda_0 e^{-\sqrt{2}\mu t} \right)}^{\frac{1}{2}}.
\ee

We note that the general solution of scale factor for a $\Lambda$CDM cosmology goes as $a(t) \sim {\left(A e^{\alpha t} + B e^{-\alpha t} \right)}^\frac{2}{3}$ where $A$, $B$, and $\alpha$ are independent constants. This may somewhat resemble equation Eq. (\ref{a_higgs}) except for the exponent value being $\frac{1}{2}$ instead of $\frac{2}{3}$ and that we have two independent parameters $\lambda_0$ and $\mu$ instead of three.

The general solution for $\psi(t)$ in this case can be written in terms of Gauss Hypergeometric function as follows

\begin{equation}  \label{psi_mod3}
\psi(t) =  \psi_0 \frac{ \frac{1}{a(t)}}{ \int \frac{1}{a(t)} \, dt} = \psi_0 \frac{\mu}{\sqrt{2} \sqrt{1-\frac{e^{2 \sqrt{2} \mu t}}{2 {\mu}^2 \lambda_0}} \, _2F_1\left(\frac{1}{4},\frac{1}{2};\frac{5}{4};\frac{e^{2 \sqrt{2} \mu  t}}{2 {\mu}^2 \lambda_0}\right)}
\end{equation}

The solution for $\psi$ is non-trivial to work with without any further simplification. As we are interested in a late time cosmological dynaimics, we note that in the limit $t \rightarrow \infty$, a hypergeometric function can be written as a simple series expansion. Morepver, for large $t$ the term containing $e^{\sqrt{2} \mu t}$ dominates over the term of $e^{-\sqrt{2} \mu t}$ in the expression of scale factor as in Eq. (\ref{a_higgs}). Thus, we can write,

\begin{eqnarray} \label{approxhiggs}
a(t) & \simeq & \frac{1}{\sqrt{2}\mu} e^{\frac{\mu t}{\sqrt{2}}}\\
\psi(t) & \simeq & D_1 \exp{(-D_2 t)}
\end{eqnarray} 
where, 
\begin{eqnarray*}
D_1 & = & \frac{2 \psi_0}{\lambda_1 \lambda_3} {\left(\frac{1}{2 {\lambda_3}^2 \mu^2} \right)}^{\frac{1}{4}} ; \\
D_2 & = & \sqrt{2} - 1/\sqrt{2} = \frac{1}{\sqrt{2}}; \\ 
\lambda_1 & = & \frac{\Gamma(\frac{5}{4})}{\Gamma(1)} ;\\
\lambda_3 & = & \frac{\Gamma(-\frac{1}{2})}{\Gamma(\frac{3}{4})} .
\end{eqnarray*}

\begin{figure} [h!]
	\begin{minipage}{\columnwidth}
		\centering
		\includegraphics[angle=0, width=0.8\textwidth]{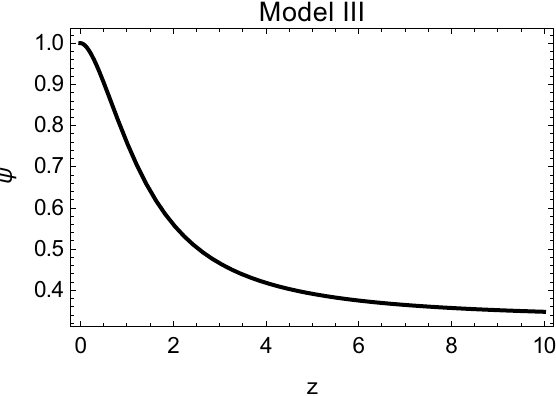}
	\end{minipage}
	\caption{{\small Plot of the Quintessence scalar field $\psi$ with redshift for Model $III$.}}
	\label{psiM3plot}
\end{figure}

\begin{figure} [h!]
\begin{minipage}{\columnwidth}
\centering
		\includegraphics[angle=0, width=0.8\textwidth]{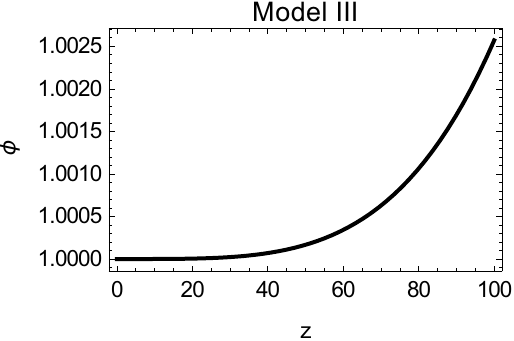}
\end{minipage}
\caption{{\small Plot of the Brans Dicke scalar field $\phi$ as a function of redshift for Model $III$, scaled with $\phi_0$.}}
	\label{phiM3plot}
\end{figure}

Using the approximate solution for $\psi$ in Eq. (\ref{evophi}) one gets a differential equation for the BD Scalar field $\phi$ which is solved numerically. The solution of the BD scalar field $\phi$, scaled with $\phi_0$ is plotted as a function of redshift in Fig. \ref{phiM3plot}. The value of BD parameter $\omega$ is again taken to be $60,000$. Moreover, the evolution of the Quintessence scalar field $\psi$ with redshift is plotted in Fig. \ref{psiM3plot}.

The nonminimally coupled scalar field theories allow for a variation of the strength of the gravitational interaction as well. As $\frac{1}{\phi}$ behaves as the effective Newtonian gravitational constant $G$ \cite{weinbook}, one can write,
$$\frac{\dot{G}}{G} = -\frac{\dot{\phi}}{\phi} = +\frac{k}{H} \mbox{ where,  $k \leq 1$ }$$
\be \label{condition}
~~~~~~~~~~~~~~~~\left\vert \frac{\dot{G}}{G}\right\vert_{z=0} \equiv \left\vert \frac{\dot{\phi}}{\phi}\right\vert_{z=0} = \frac{k}{ H_0} \leq 10^{-10} \mbox{ per year.}
\ee

The values of model parameters have been taken from Table \ref{resulttable_M3}. The initial conditions for solving the second order differential equation of $\phi$ for the Models $II$ and $III$ are chosen obeying the constraint on $G$ according to Eq. (\ref{condition}).  \\

Comparing Fig. \ref{phiM3plot} and Fig. \ref{psiM3plot} we comment that at present epoch $z \sim 0$ the BD scalar field $\phi$ almost becomes a constant, while the quintessence scalar field $\psi$ has a sharp increasing behavior. At an early epoch $z >> 0$ this behavior simply reverses. One can interpret that within the domain of early universe cosmology, the BD scalar field plays a much more dominating role over the quintessence field. During present epoch, the role reverses and the quintessence field dominates the accelerated expansion. However, detailed comments on such aspects require further analysis and shall be addressed elsewhere.  \\

Using the solutions as in Eq. (\ref{approxhiggs}) one can rewrite $\psi$ and $\dot{\psi}$ as,
\begin{eqnarray} \label{psi_psidot}
\psi & = & D_1e^{-\frac{t}{\sqrt{2}}}  =  C[a(t)]^{-\mu}, \\
\dot{\psi} & = & -\frac{D_1}{\sqrt{2}}e^{-\frac{t}{\sqrt{2}}}  = -\frac{\psi}{\sqrt{2}}.
\end{eqnarray}

Thus, the expression for quintessence energy density $\rho_{\psi}$ becomes
\be \label{rhopsi}
\rho_{\psi} = V_0 + \frac{2\mu^2 +1}{4}\psi^2 + \frac{\lambda_0}{4}\psi^4.
\ee

In BD theory the scalar field is proportional to inverse of the Gravitational constant. Therefore one can write
\be \label{phihiggs}
\phi = \phi_0[a(t)]^{\epsilon} ~~\mbox{ such that }~~ \dot{\phi} = \epsilon H \phi.
\ee 

Using equations (\ref{psi_psidot}), (\ref{rhopsi}) and (\ref{phihiggs}) in the Friedmann equation (\ref{EE0}), we get the expression for reduced Hubble parameter as,
\be \label{hubble_higgs}
h^2 = \frac{\Omega_{m0} a^{-(3+\epsilon)} + \frac{V_0}{3H_0^2 \phi_0}a^{-\epsilon} +  E_1 a^{-(2\mu+\epsilon)} +  E_2 a^{-(4\mu+\epsilon)}}{ \left[1-\frac{\omega}{6}\epsilon^2 +\epsilon \right]}
\ee
where,
\begin{eqnarray*}
E_1 & = & \frac{C^2 (2\mu^2+1)}{12 H_0^2 \phi_0}\\
E_2 & = & \frac{C^4 \lambda_0}{12 H_0^2 \phi_0}
\end{eqnarray*}

Figure \ref{phiM3plot} depicts that in a late time scenario, $\phi$ behaves as a constant ($\phi_0$), which implies $\epsilon$ can be considered $0$ in present epoch. So, the modified expression for the reduced Hubble parameter becomes,
\be
h^2 = \Omega_{m0} a^{-3} + \frac{V_0}{3H_0^2 \phi_0} +  E_1 a^{-2\mu} +  E_2 a^{-4\mu}.
\ee

Using the constraint equation such that, in the present epoch at $z = 0$, $a = a_0 = 1$, $h = 1$, we are finally left with,
\be \label{finalmodel}
h^2 = \Omega_{m0} a^{-3} +  E_1 a^{-2\mu} +  E_2 a^{-4\mu} + (1-\Omega_{m0}-E_1-E_2)
\ee

Note that, if $\mu$ is chosen as unity, equation (\ref{finalmodel}) exactly mimics a $\Lambda CDM$ model where, the constraint term $\frac{V_0}{3H_0^2 \phi_0}$ behaves as Cosmological Constant.

\subsection{{\bf Brans Dicke with a Quintessence and a general combination of Power Law Potentials (Model $IV$)}}
We also present an example where the self-interaction potential of the quintessence scalar field is a simple combination of power functions $\sim$ $\psi^2+ \psi^\delta$, not restricting $\delta$ to $4$ only. Similar to the higgs potential, this allows us to observe the role of a quadratic term in the potential. Moreover, this allows one more parameter in the solutions, which is the exponent, allowing wider variety of solutions and possibilities. The potential can be written as

\be
V(\psi) = \frac{1}{2}\psi^2 + \frac{\psi^{(n+1)}}{(n+1)},
\ee

using which one can write the scalar field evolution Eq. (\ref{evopsi}) as
\be \label{psi_combo}
\ddot{\psi} + 3\frac{\dot{a}}{a}\dot{\psi}+ \psi + \psi^n = 0.
\ee

On a comparison, this gives a class of the general oscillator equations (\ref{anharmonic}) for different $n$. Using the aforementioned method of integrability we arrive at the evolution equation for scale factor $a(t)$ and Quintessence field $\psi(t)$ written as

\begin{eqnarray}
&& a(t) = {\left[ \delta_0 \cosh{\sqrt{\frac{(n+1)}{2}} \left\lbrace t +6\left(3+n\right)\delta \right\rbrace} \right]}^{\frac{(n+3)}{3 \left(n+1 \right)}}, \label{acombo} \\&& 
\psi(t) = -\frac{2n\sqrt{1-y(t)}y(t)}{3\sqrt{(1+n)}z(t)}, \\&& \nonumber
y(t) = \cosh\left[ \sqrt{2(1+n)}\left\lbrace t+6(3+n)\right\rbrace \right], \\&& \nonumber
z(t) = _{2}F_{1} \left[\frac{1}{2},\frac{n}{3(n+1)};\left(\frac{3+4n}{3+3n}\right) ; z_{1}(t) \right], \\&& \nonumber
z_{1}(t) = \left\{ \cosh \left(\sqrt{\frac{1+n}{2}} ({t+6(3+n)\delta_1})\right) \right\}^2.
\end{eqnarray}

Carrying out a similar treatent as in Section $3.1$ one may find that the qualitative behavior of the Quintessence scalar field $\psi$ and the BD scalar field $\phi$ are similar to that of Model $III$, apart from some scaling.

\section{Comparison with Observational Data for the Late time accelerating cosmologies}
\subsection{{\bf Parameter Estimation}}
The Models in section $3$, i.e., models in $III$ and $IV$, involving a combination of two potential functions, gives the deceleration parameter as a function of redshift, which might show a possibility for signature flip in the expression of $q$. Therefore, it is prescribed to estimate the model parameters and study the evolution of the different cosmological quantities extensively for Model $III$ and $IV$. In this section, we estimate the model parameters and study the confidence contours on the parameter space, the marginalized likelihood function using four different data sets; the Supernova distance modulus data (SNe), observational measurements of Hubble parameter (OHD), Baryon Acoustic Oscillation (BAO) data and the Cosmic Microwave Background (CMB) data. 
\\

We first write the cosmological quantities in a dimensionless way from the expression of scale factor $a(t)$ and it's time derivatives, using the redshift as the argument instead of cosmic time $t$. Redshift $z$ is a dimensionless observational quantity defined as,
\be \label{redshift}
(1 + z) = \frac{a_0}{a(t)}, 
\ee
$a_0$ being the present value of the scale factor. The Hubble parameter $H(t) = \frac{\dot{a}}{a}$ can also be written as a function of the redshift $z$ as

\be
H(z) = -\frac{1}{(1 + z)} \frac{dz}{dt}.
\ee

We can estimate the Hubble parameter from equation (\ref{a_higgs}) and (\ref{acombo}) by rewriting it as a perfect square term. Using the method of substitution and writing down the exponential term as functions of redshift, we get the parametric expression for the Hubble parameter as

\begin{eqnarray*}
H(z)_{III} & = & H_0\sqrt{\lambda_0 {(1+z)}^4 + \frac{{\mu}^2}{2}}\\
H(z)_{IV} & = & \sqrt{\frac{(n+3)^2}{18(n+1)}\left(1-H_0^2 (1+z)^{\frac{6(n+1)}{n+3}} \right)+ H_0^2 (1+z)^{\frac{6(n+1)}{n+3}} }
\end{eqnarray*}
where we scale the present value of Hubble parameter, $H_0$ by $100$ km $\mbox{Mpc}^{-1}$ $\mbox{sec}^{-1}$ and represent it in a dimensionless form as $h_0$. Thus, the reduced Hubble parameter becomes,
\be
h(z) = \frac{H(z)}{H_0} = \frac{H(z)}{100\times h_0}
\ee

We note that the contribution of the Higgs field towards the total energy density of the Universe resembles a radiation dominated Universe in the early epoch. \\
For parameter estimation, we use the distance modulus measurements of type Ia supernova from the Joint Light-Curve Analysis following the work of Betoule et. al. \cite{jla}, who studied cosmological constraints from the SN-Ia observations of SDSS-II and SNLS collaborations. We incorporate the estimation of the Hubble parameter as a function of redshift \cite{ohd}, alongwith the measurement of Hubble parameter from Lyman-$\alpha$ forest at redshift $z=2.34$ by Delubac et. al. \cite{delu} and measurement of $H_0$ from Planck \cite{planck}. For BAO data, three independent measurements of $\frac{r_s(z_d)}{D_v(z_{BAO})}$ are used. $r_s(z_d)$ gives the sound horizon at photon drag epoch ($z_d$) and $D_v$ is the dilation scale at the redshift of BAO measurement. Three measurements are for three different values of redshift, for instance, from $6dF$ $Galaxy$ $Survey$ at $z = 0.106$ \cite{6dF}, from Baryon Oscillation Spectroscopic Survey (BOSS) at $z = 0.32$ (BOSS LOWZ) and at $z = 0.57$ (BOSS CMASS) \cite{bossanderson}. The BAO measurements have been scaled by the acoustic scale ($l_A$) estimated from Planck \cite{planck}. The CMB shift parameter is related to the position of the first acoustic peak in power spectrum of the temperature anisotropy of the CMB radiation. Value of the parameter is estimated from the CMB data along with some assumption about the model of background cosmology, as estimated from Planck data \cite{planck}.

\begin{figure} [h!]
\begin{minipage}{\columnwidth}
`	\centering
		\includegraphics[angle=0, width=0.95\textwidth]{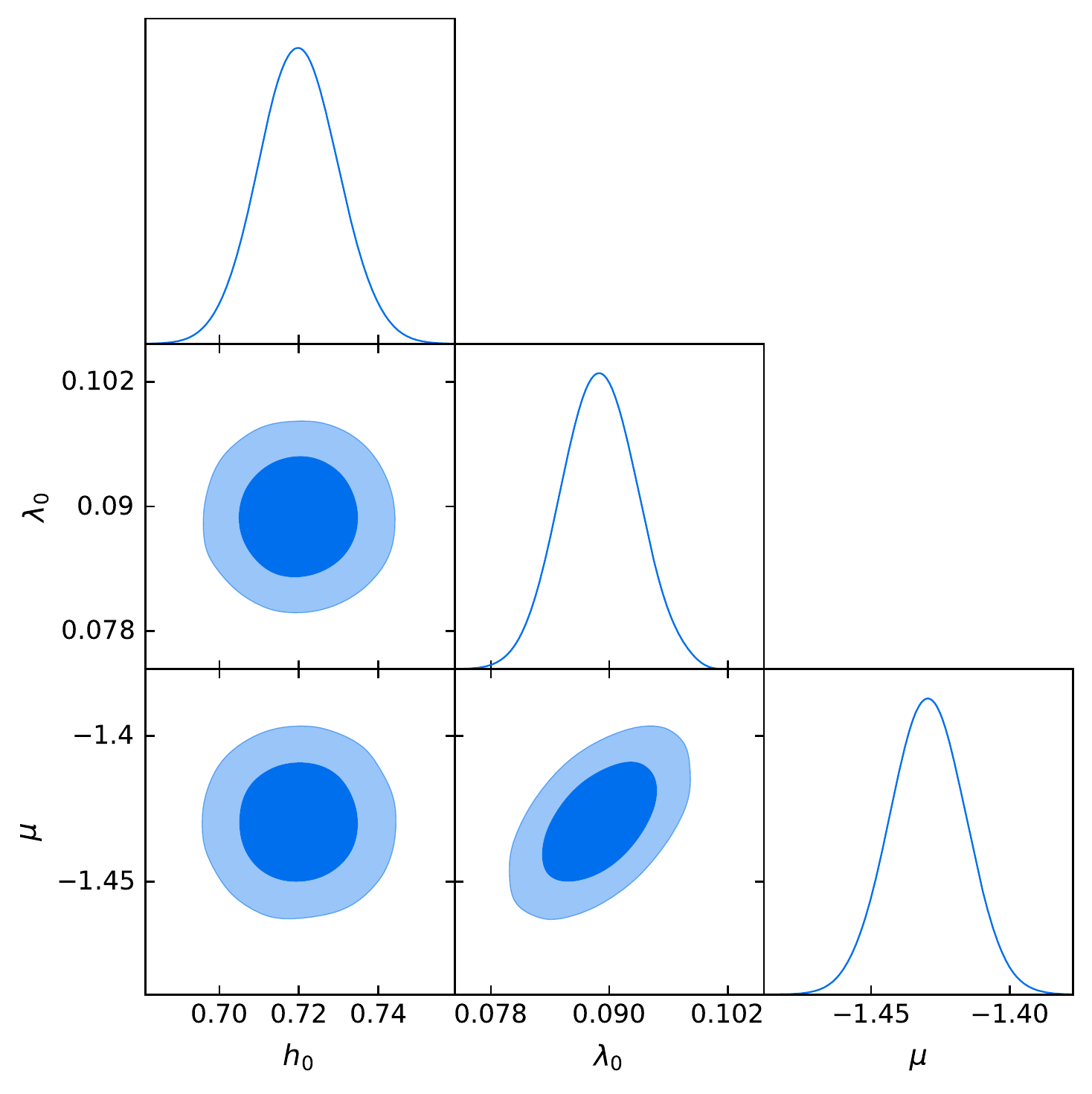}
		\includegraphics[angle=0, width=0.7\textwidth]{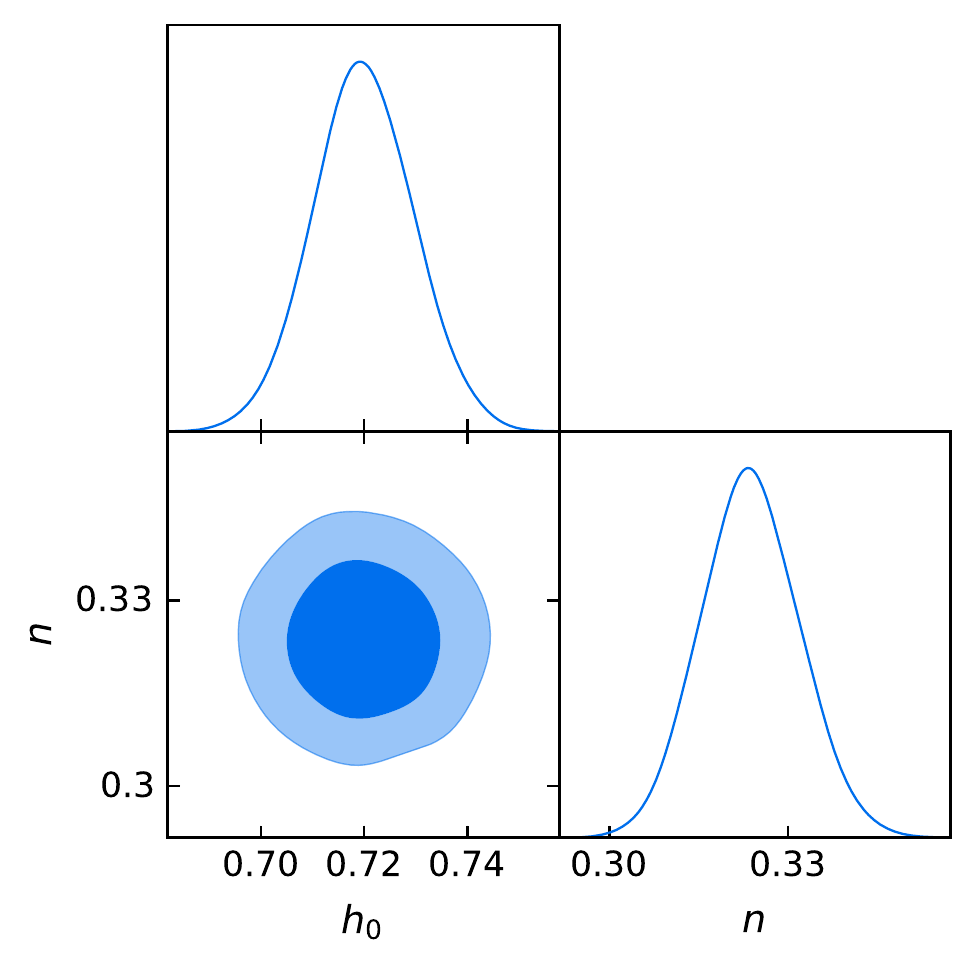}
\end{minipage}
\caption{{\small (i) Confidence contours on the parameter space and the marginalized likelihood function of Model $III$, obtained from the combined analysis of OHD+JLA+BAO+CMB are shown on the figure on top. The associated 1$\sigma$, 2$\sigma$  confidence contours are shown. (ii) Confidence contours on the parameter space and the marginalized likelihood function of Model $IV$, obtained from the combined analysis of OHD+JLA+BAO+CMB are shown on the figure below. The associated 1$\sigma$, 2$\sigma$ confidence contours are shown.}}
\label{Modelcontour}
\end{figure}

\begin{table}
\caption{{\small The parameter values and the associated 1$\sigma$ uncertainty of the parameters of Model $III$, obtained from the analysis with different combinations of the data sets.}}\label{resulttable_M3}
\begin{tabular*}{\columnwidth}{@{\extracolsep{\fill}}lrrrrl@{}}
\hline
 & \multicolumn{1}{c}{$h_0$} & \multicolumn{1}{c}{$\lambda_0$} & \multicolumn{1}{c}{$\mu$} \\
\hline
$OHD+JLA$		  & $0.720^{+0.011}_{-0.010}$ & $0.089^{+0.004}_{-0.004}$ & $-1.429^{+0.014}_{+0.014}$ &\\
\\
$JLA+BAO$		  & $0.720^{+0.010}_{-0.010}$ & $0.034^{+0.003}_{-0.002}$ & $-1.481^{+0.012}_{-0.012}$ &\\
\\
$OHD+JLA+BAO$ 	  & $0.720^{+0.010}_{-0.010}$ &$0.065^{+0.003}_{-0.003}$ & $-1.496^{+0.013}_{-0.012}$ &\\
\\
$OHD+JLA+BAO+CMB$ & $0.719^{+0.010}_{-0.009}$ &$0.042^{+0.003}_{-0.003}$ & $-1.517^{+0.012}_{-0.011}$ &\\
\hline
\end{tabular*}
\end{table}

\begin{table}
	\caption{{\small The parameter values and the associated 1$\sigma$ uncertainty of the parameters of Model $IV$, obtained from the analysis with different combinations of the data sets.}}\label{resulttable_M4}
	\begin{tabular*}{\columnwidth}{@{\extracolsep{\fill}}lrrrrl@{}}
		\hline
		&  \multicolumn{1}{c}{$h_0$} &   &   &   \multicolumn{1}{c}{$n$} &\\
		\hline
		$OHD+JLA$		  &  $0.719^{+0.009}_{-0.010}$ &   &   &  $0.264^{+0.021}_{-0.020}$ & \\
		\\
		$JLA+BAO$		  &  $0.720^{+0.010}_{-0.010}$ &   &   &  $0.565^{+0.017}_{-0.017}$ & \\
		\\
		$OHD+JLA+BAO$     &  $0.719^{+0.009}_{-0.010}$ &   &   &  $0.380^{+0.009}_{-0.009}$ & \\
		\\
		$OHD+JLA+BAO+CMB$ &  $0.719^{+0.009}_{-0.009}$ &   &   &  $0.326^{+0.008}_{-0.008}$ & \\
		\hline
	\end{tabular*}
\end{table}

Uncertainty of the parameters are estimated by `Markov Chain Monte Carlo' (MCMC) method with the assumption of a uniform prior distribution. In the present analysis, we have adopted a Python implementation of the ensemble sampler for MCMC, the `emcee', introduced by Foreman-Mackey {\it et al.} \cite{emcee}.
\\
Fig. \ref{Modelcontour} shows the confidence contours on the parameter space and the marginalized likelihood function of Model $III$ and $IV$ obtained from the combined analysis with different datasets. It clearly depicts that, the parameters $h_0$ and $\lambda_0$ for Model $III$ have a positive correlation between themselves. The best-fit values of the model parameters have been estimated for both cases, and the associated 1$\sigma$ uncertainty, obtained for different combinations of the data sets in case of Model $III$ and $IV$ are represented in Table \ref{resulttable_M3} and \ref{resulttable_M4} respectively.

\subsection{{\bf Evolution of Cosmological Parameters}}

\begin{figure} [h!]
\begin{minipage}{\columnwidth}
`	\centering
		\includegraphics[angle=0, width=0.8\textwidth]{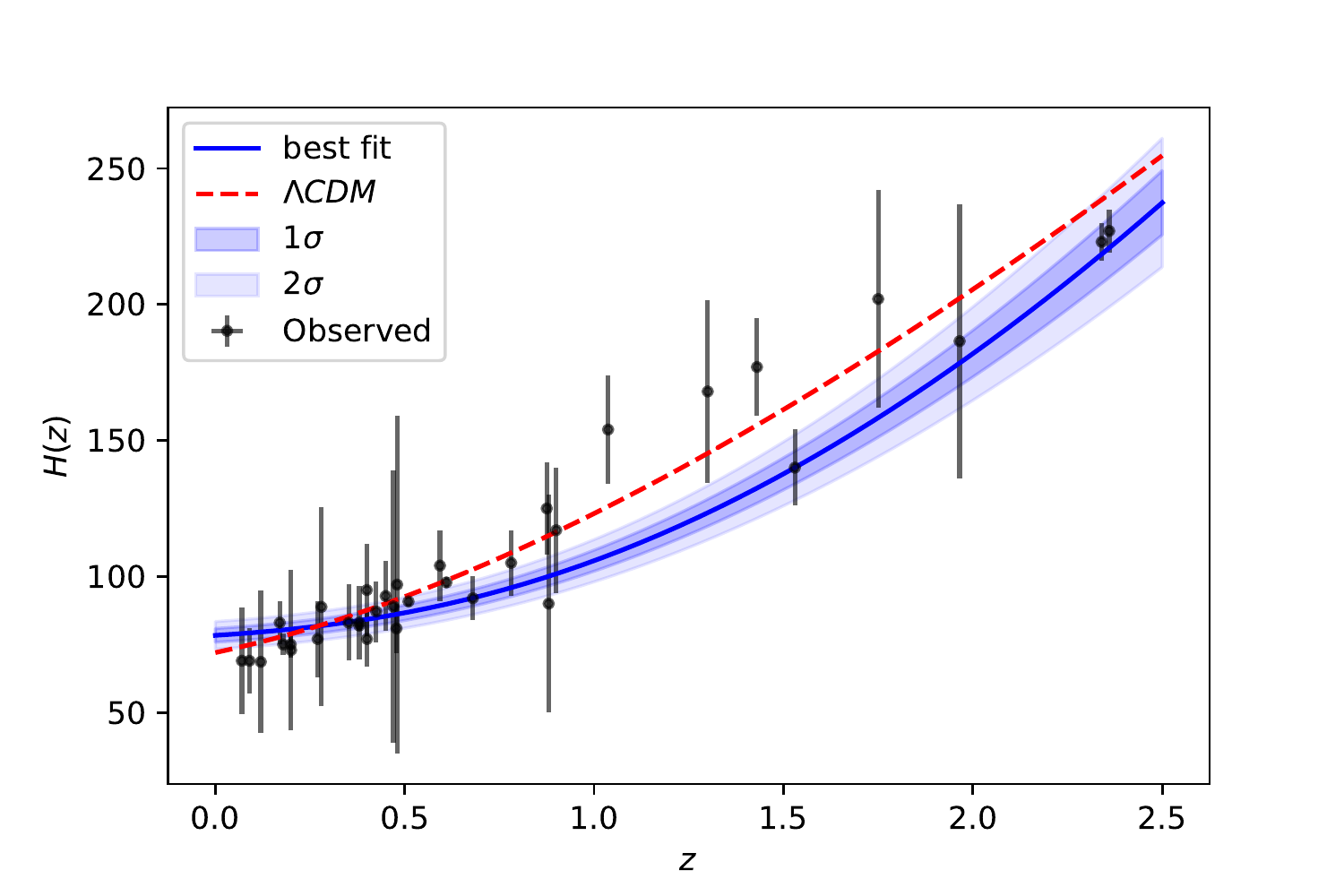}
\end{minipage}
	\caption{{\small Plot of the reconstructed Hubble parameter $H(z)$ for Model $III$, Higgs Potential. The best fit values and the associated 1$\sigma$, 2$\sigma$ confidence regions are obtained from the combined analysis.}}
	\label{Hzplot_higgs}
\end{figure}

\begin{figure} [h!]
\begin{minipage}{\columnwidth}
`	\centering
		\includegraphics[angle=0, width=0.8\textwidth]{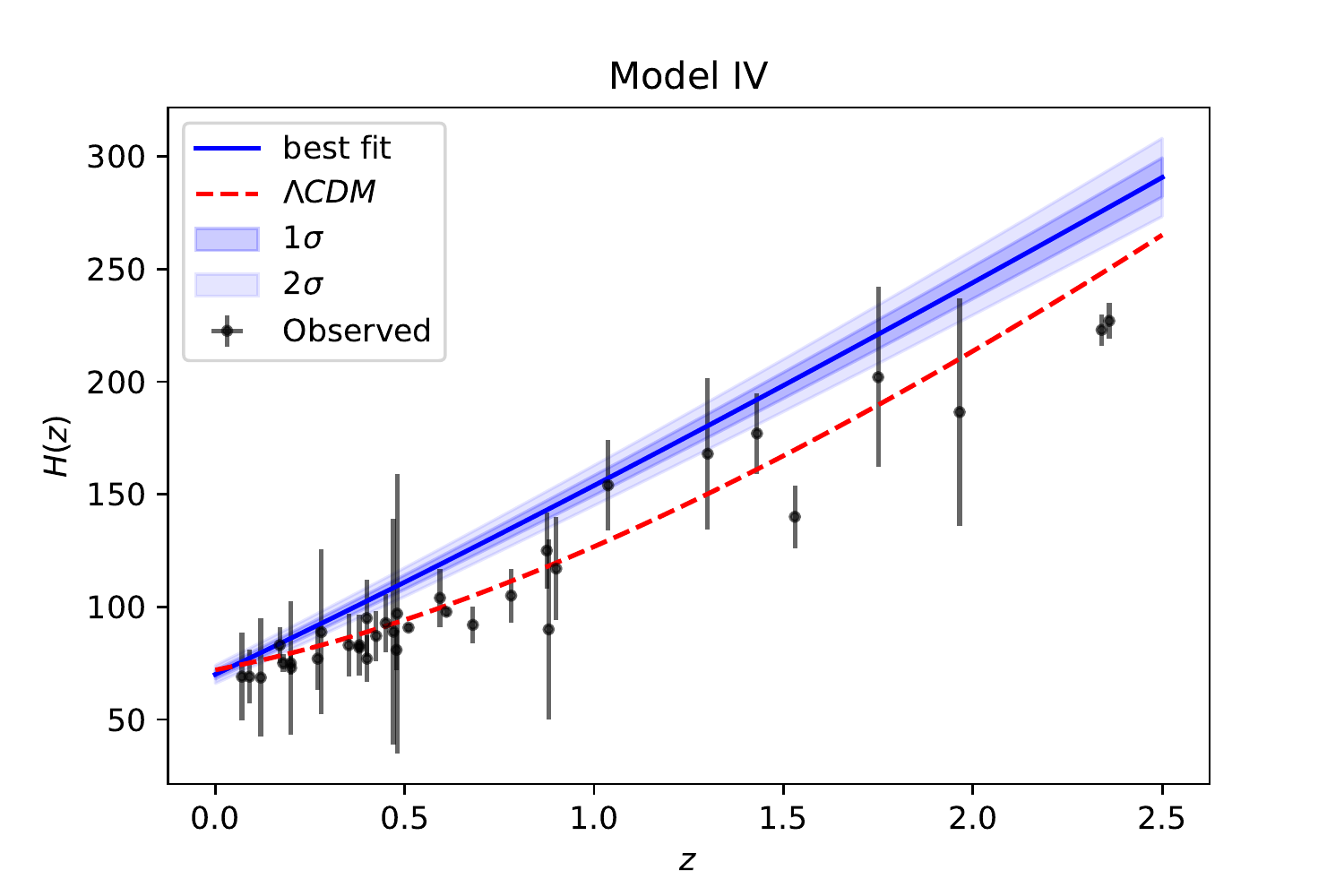}
		\includegraphics[angle=0, width=0.8\textwidth]{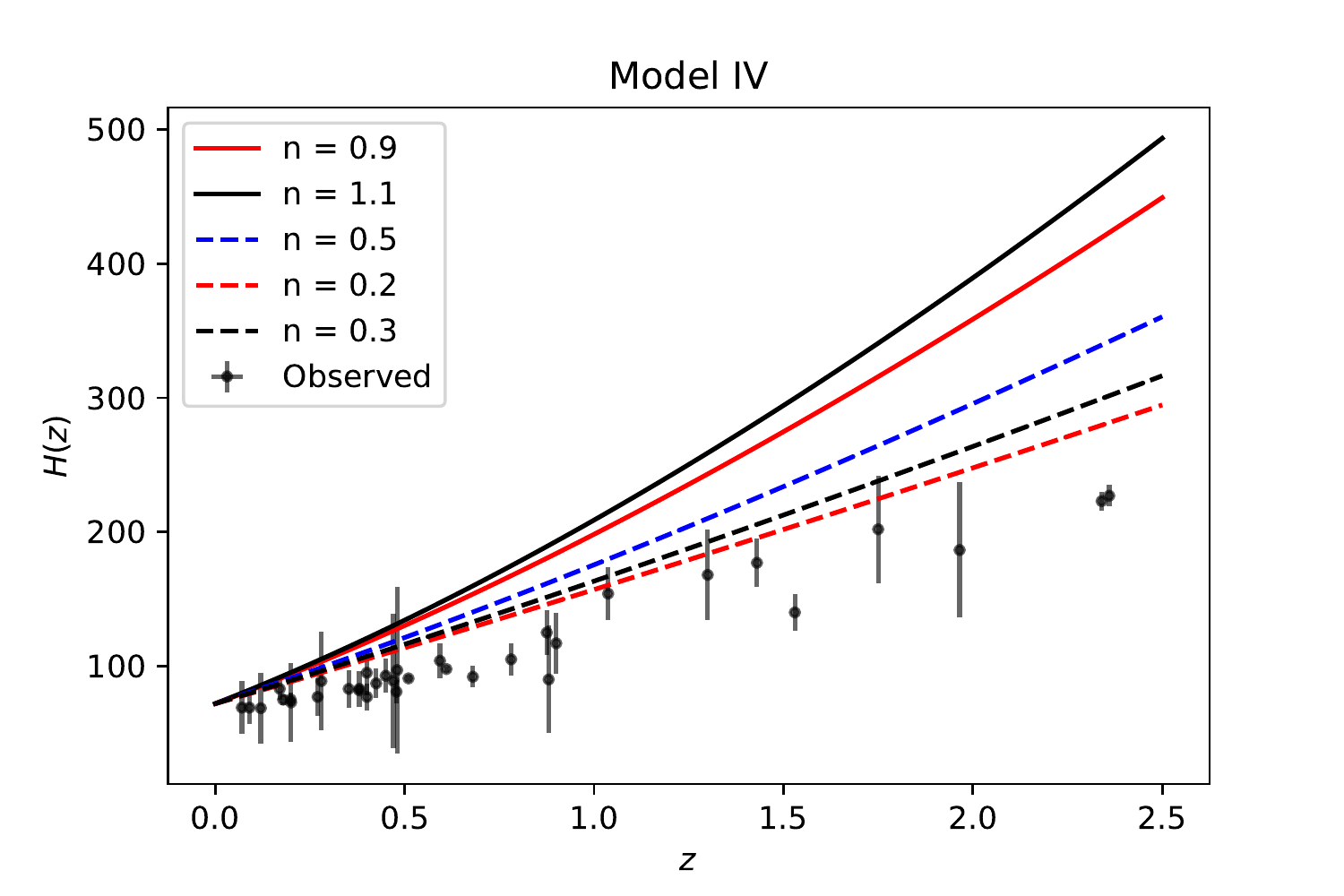}
\end{minipage}
	\caption{{\small Plots of the reconstructed Hubble parameter $H(z)$ for Model $IV$, general combination of power law potential. The best fit values and the associated 1$\sigma$, 2$\sigma$ confidence regions are obtained from the combined analysis. Hubble parameter plots using different values of $n$ for Model $IV$ are also shown}}
	\label{Hzplot_general}
\end{figure}

Depending on the best fit choice of model parameters, the functional form of the cosmological quantities can be plotted as a function of redshift. Figs. \ref{Hzplot_higgs} and \ref{Hzplot_general} show that the plots of $H(z)$ for the reconstructed models $III$ and $IV$ are consistent with the observational data in the low redshift regime. As the value of $z$ increases ($z > 1.0$) a discrepancy arises between the theoretical and observational results. Note that the best fit value for Hubble parameter at present epoch, i.e, $H_0$ is close to the result obtained by Riess {\it et al.} \cite{riess2018}.\\

The kinematic quantities related to the expansion of our universe play a vital role in cosmology. They basically are connected to the second and third order derivatives of the scale factor. For instance, the deceleration parameter is written as,
\be \label{deceleration}
q = - \frac{\ddot{a}}{aH^2(t)}  \equiv -1 + \frac{1}{2} (1+z) \frac{{[H^2(z)]}^{'}}{H^2(z)}.
\ee

The jerk parameter can be written as
\be \label{jerk}
j = \frac{\dddot{a}}{aH^3(t)} \equiv  1 - (1+z)\frac{{[H^2(z)]}^{'}}{H^2(z)} + \frac{1}{2}{(1+z)}^2 \frac{{[H^2(z)]}^{''}}{H^2(z)}.
\ee

The effective equation of state parameter $w_{eff}$, represented by a ratio of the total pressure contribution to the total energy density in our universe, can be written as a function of $z$ as
\be 
w_{eff}(z) =  - 1 + \frac{2}{3} (z+1) \frac{H'(z)}{H(z)}. 
\ee


\begin{figure} [h!]
\begin{minipage}{\columnwidth}
`	\centering
		\includegraphics[angle=0, width=0.8\textwidth]{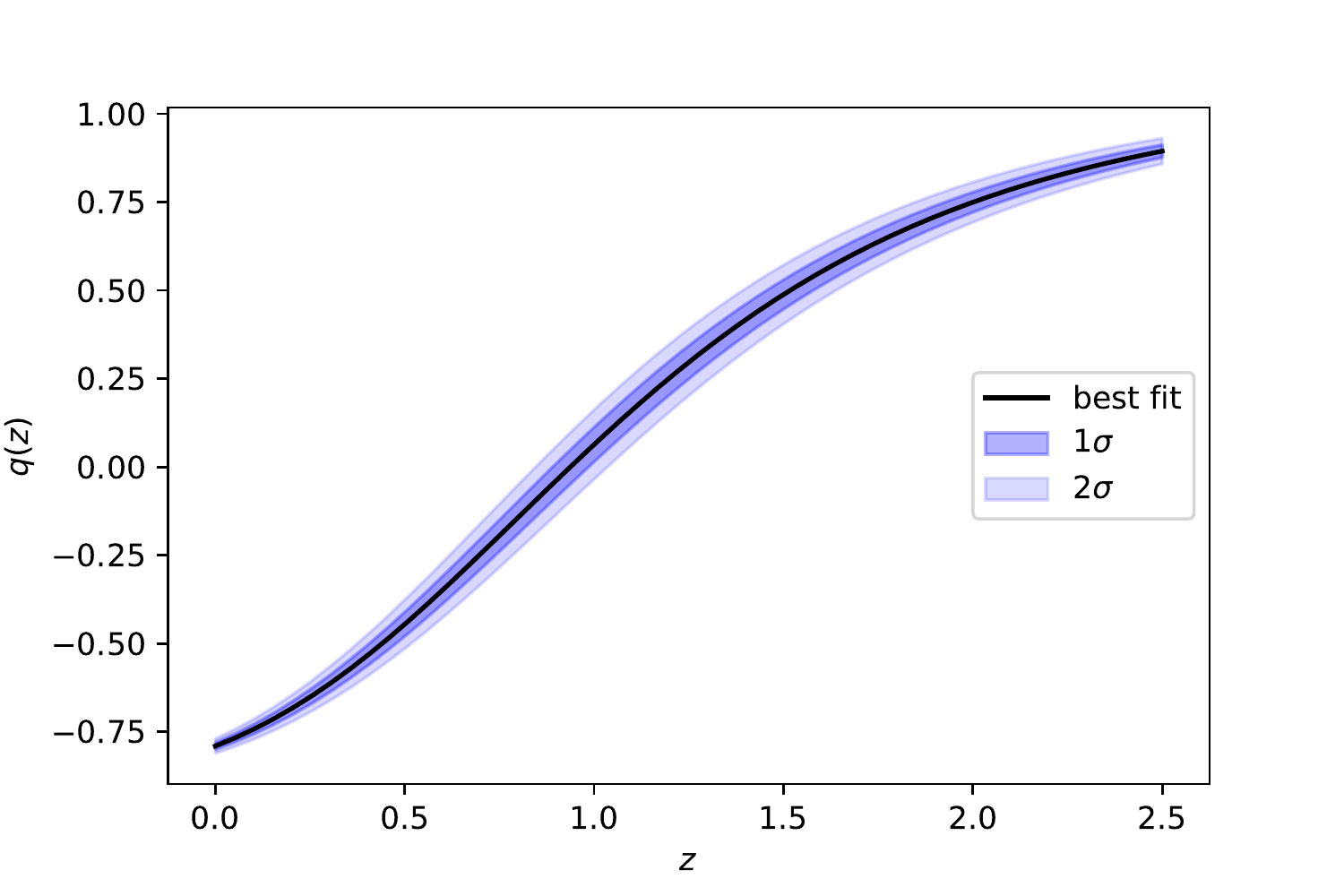} 
\end{minipage}
	\caption{{\small The deceleration parameter $q(z)$ plot for Model $III$ using the best fit values and the associated 1$\sigma$, 2$\sigma$ confidence regions from the combined analysis.}}
	\label{qzplot}
\begin{minipage}{\columnwidth}
`	\centering
		\includegraphics[angle=0, width=0.8\textwidth]{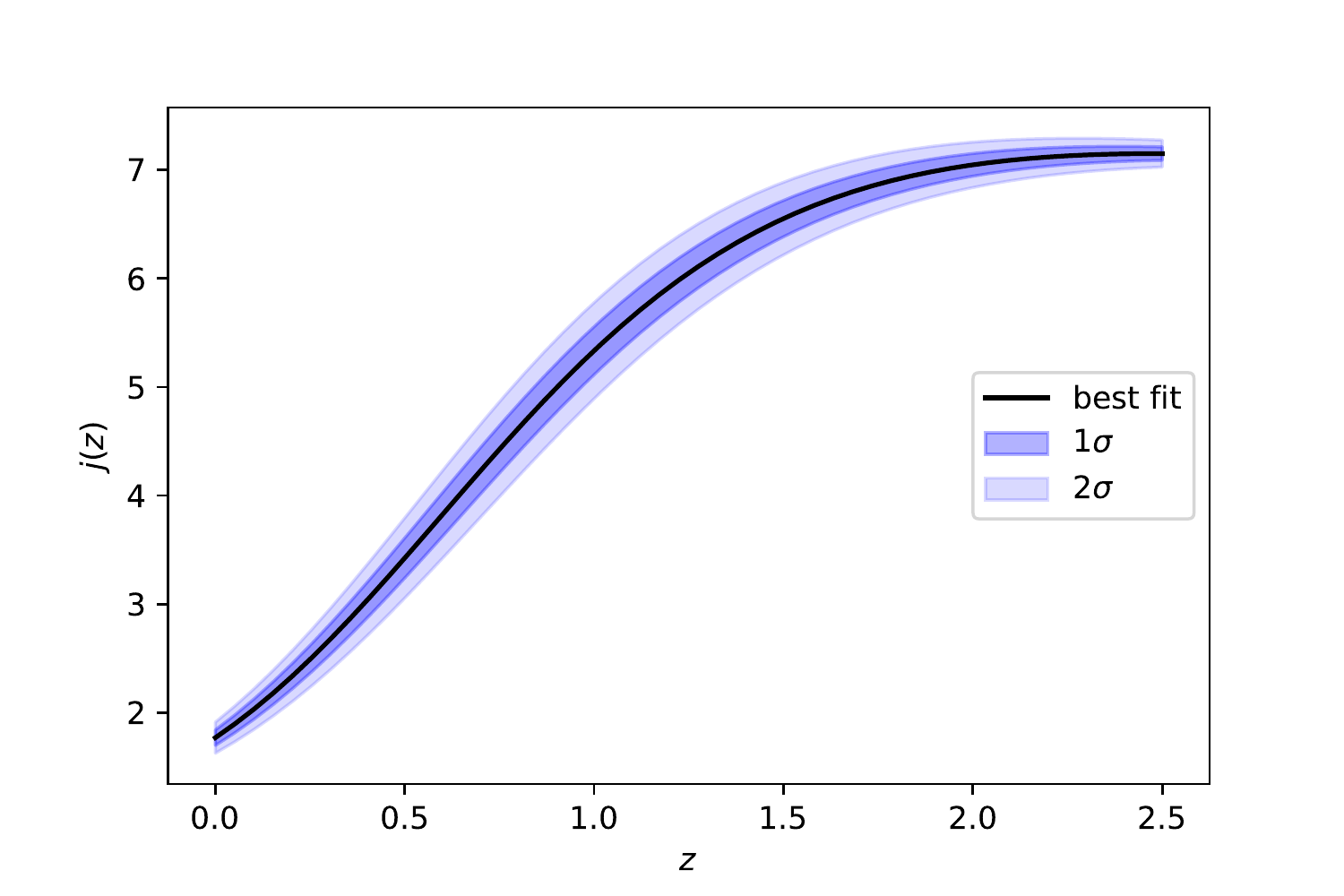} 
\end{minipage}
	\caption{{\small The jerk parameter $j(z)$ plot for Model $III$ using the best fit values and the associated 1$\sigma$, 2$\sigma$ confidence regions from the combined analysis.}}
	\label{jzplot}
\end{figure}

Fig. \ref{qzplot} shows that for Model $III$, there is a transition in the signature of $q(z)$ from a decelerated phase to an accelerated phase of expansion. Moreover, the transition redshift $z_{t} < 1$ is consistent with direct observational results \cite{riess2004, fqrat}. Thus, it can be said that the Higgs' interaction potential model is consistent with the observed evolution of $q(z)$. But in case of Model $IV$ no such transition behavior could be seen. So, Model $IV$ is inconsistent with the observed evolution of $q$, and hence can be ruled out. This situation probably arises because of the fact that in Model $IV$ the coefficients for combined potential functions is considered as unity, which hints towards a uncompensated reduction in the necessary model parameters. \\

\begin{figure} [h!]
\begin{minipage}{\columnwidth}
`	\centering
		\includegraphics[angle=0, width=0.8\textwidth]{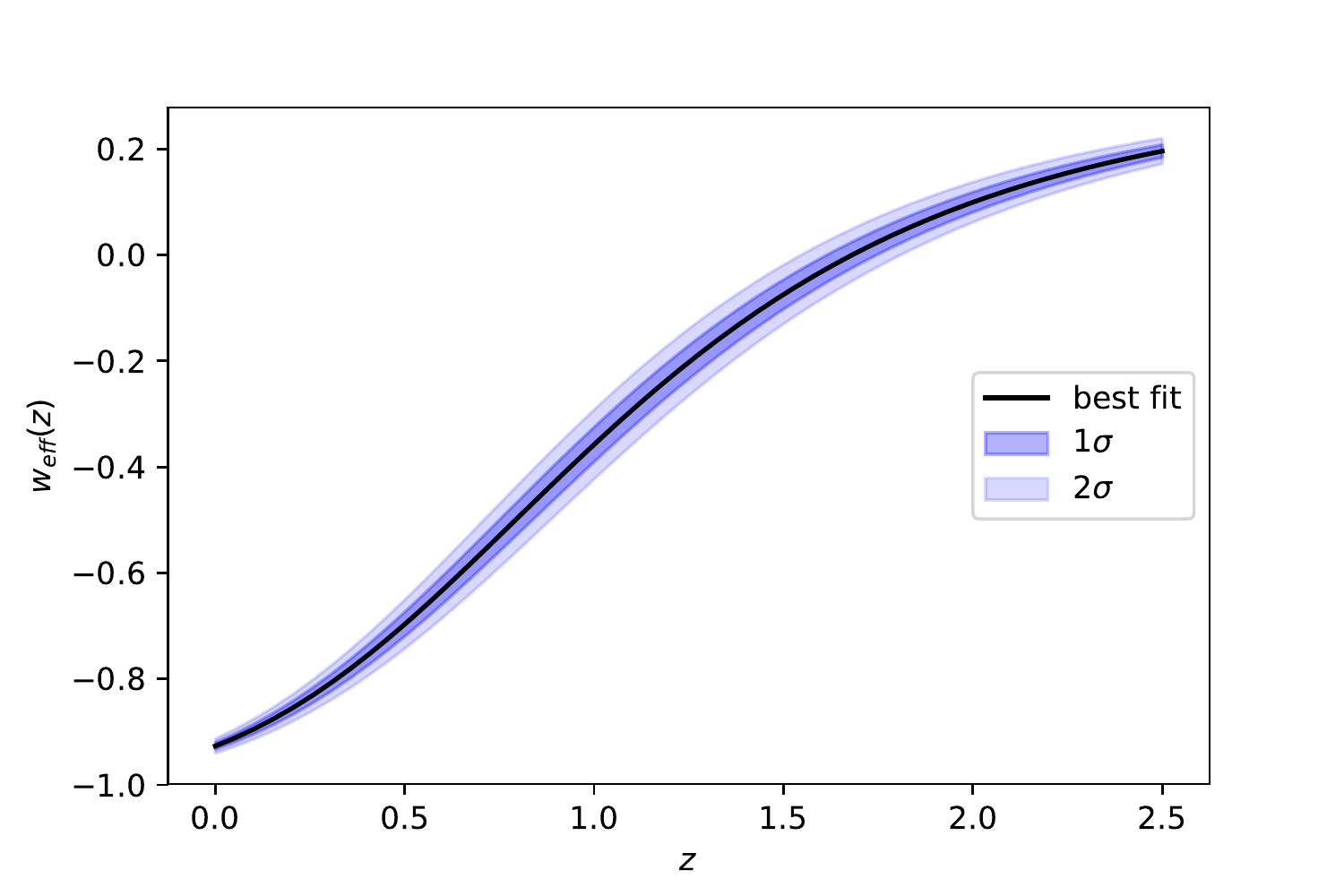} 
		\includegraphics[angle=0, width=0.8\textwidth]{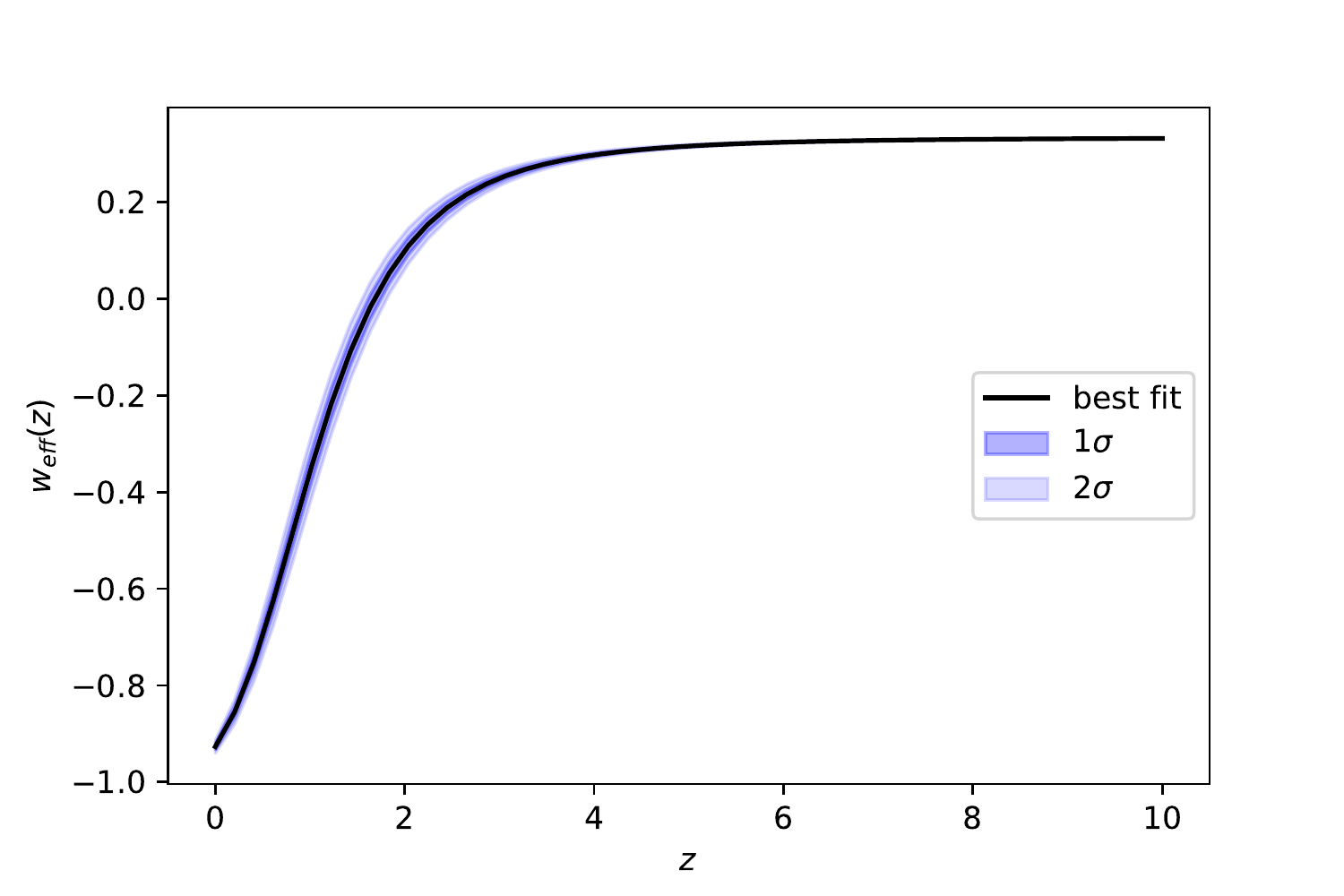} 
\end{minipage}
	\caption{{\small Plots for the effective equation of state of Model $III$ using different ranges of redshift. The best fit, 1$\sigma$ and 2$\sigma$ confidence regions, obtained in the combined analysis are shown.}}
	\label{ww}
\end{figure}

The deceleration parameter is expected to show a non-trivial evolution with respect to redshift as recent observational cosmology suggests. Therefore it is very important to investigate the next order derivative of the scale factor or the jerk parameter, whose value is unity for $\Lambda$CDM model. Model $III$ shows a non-trivial evolution of jerk parameter as a function of redshift, shown in Fig. \ref{jzplot}. The present value of jerk parameter obtained from Model $III$ remain in between $0.9$ to $1.3$ at 1$\sigma$ level, well in agreement with the requirement of observational evidences. This also confirms the recent prediction that jerk parameter is in general expected to be in close proximity with the corresponding $\Lambda$CDM value \cite{ammnras}. \\

The graph on the top of Fig. \ref{ww} shows the evolution of $w_{eff}$ for Model $III$. At the present epoch, $z = 0$, $w_{eff}$ is around $-1$ but slightly greater. While this may indicate a `phantom' nature, it also strongly indicates a $\Lambda$CDM behavior depending on different model parameter values. With increase in $z$, $w_{eff}$ gradually rises and becomes a constant at $\sim 0.3$ which resembles a radiation dominated universe at high redshift (bottom graph of Fig. \ref{ww}).

\begin{table}
	\caption{{\small Comparison of the cosmographical parameters}}\label{res_cosmo}
	\begin{tabular*}{\columnwidth}{@{\extracolsep{\fill}}lrrrrl@{}}
		\hline
		& \multicolumn{1}{c}{$H_0$} & \multicolumn{1}{c}{$q_0$} & \multicolumn{1}{c}{$j_0$} \\
		\hline
		$Higgs$		  & $72.34$ & $ -0.79 $ & $1.16$ &\\
		\\
		$\Lambda CDM$		  & $70.46$ & $-0.61$ & $1.0$ &\\
		\\
		$Observed$ 	  & $69.8 \pm 1.9 $ \cite{hstH0} &$-0.60 \pm 0.2$ \cite{planck}& $ $ &\\
		\hline
	\end{tabular*}
\end{table}

In Table \ref{res_cosmo}, we compare different values cosmographical parameters for the Brans-Dicke + Higgs (Model $III$) setup with corresponding values of the parameters from $\Lambda$CDM and observations.

\section{Conclusion} \label{dis}

It can not be denied that Brans-Dicke theory might have renounced it's original appeal a little bit as long as it can not reproduce GR at some limit of the BD parameter $\omega$. However, this does not diminish the stature of the theory as a prototype of scalar-tensor theories of gravity. Moreover, the theory has the elegance of describing both the inflationary universe and the present accelerated expansion of the universe without any need of dissipative processes or an exotic fluid. One often considers generalizations or modifications of the theory in a hope of tackling the shortcomings and making it a `better' theory of gravity. \\

The present work deals with accelerating solutions in modifed BD theory. Keeping in mind the non-linearity of the system of field equations, a mathematical method of treating an anharmonic oscillator equation system is incorporated. The BD scalar field evolution equation is treated with an Euler-Duarte-Moreira method of integration. The advantage of this method is that, it helps one to solve the system of equations without any apriori assumption on the scale factor or the scalar field from the cosmic history, but rather, the restriction over the choice of the functions come from a purely mathematical property. \\

The solutions for three different models are discussed. For model $III$, a parameter estimation is carried out and the confidence contours on the parameter space are studied using four data sets, namely, SNe, OHD, BAO and the CMB data. The cosmological quantities, for instance the effective equation of state parameter $\omega_{eff}$, the deceleration and the jerk parameters are plotted as a function of redshift for the best fit choices of the model parameters. For Model $III$ the evolution of $\omega_{eff}$ closely resembles a $\Lambda$CDM behavior around $z \sim 0$ and a matter dominated universe at high redshift. Model $III$ also shows a transition in the signature of $q(z)$ and the transition redshift is consistent with direct observational results. Models $I$ and $II$ has zero value of the jerk parameter whereas Model $III$ shows a non-trivial evolution of $j(z)$ with the present value of jerk parameter remaining in between $0.9$ to $1.3$ in the present epoch. We can therefore note that the Brans Dicke with quintessence or a `BDQ' setup with a Higgs interaction potential case is well consistent with the observed evolution of cosmological quantities, as compared to the other significant options. A general combination of power functions of the quintessence field is also considered as potential. In such a case, the scale factor is seen to describe an accelerated expansion. For some choices of the potential, the hubble parameter also follows closely the observational data-points. However, not all combinations produce a result consistent with observational prediction as they fail to describe the signature flip in the evolution of deceleration parameter.   \\

The simplification of the BD field equations and a subsequent extraction of an exact solution is extremely non-trivial. This difficulty is often by-passed by studying simplified systems who do not fail to describe the physics involved. In that sense, the present work also presents some special cases, as the solutions come under a specific assumption over the integrabilty of one particular equation only. However, the cosmological solutions found through the present method seem to describe the accelerated expansion of the universe quite well, at least for some specific cases. The solutions are simple and easy to work with for further allied investigations. Moreover, the simplicity provides one with the opportunity to study the evolution of the BD scalar field in a general manner. \\

It must also be mentioned that the issue regarding the value of BD parameter $\omega$ remains to be solved such that the theory can solve cosmological requirements along with satisfying the lower limit on the BD parameter, imposed by the solar system experiments. To conclude we note that different single or multiple scalar field models remain extremely popular even in the present context candidates to fill in for the fluid responsible for the late-time acceleration. However, it is always better to have a complete mathematical theory providing a unified description of the whole expansion history of the universe, from an early inflationary epoch to a late time cosmic acceleration, and beyond. It was quite extensively discussed by Elizalde et. al. \cite{elizaldereconstruction} that given a certain universe expansion history, one can reconstruct a wide class of minimally or non-minimally coupled scalar field theories presenting a number of explicit examples which show a unified description of the inflationary era and a late-time cosmic acceleration epoch. While the models discussed in the present work do not describe a unified cosmic history in that sense, it shows potential in producing interesting exact solutions describing atleast some patches of our universe, namely the late-time era. Some simple power law solutions also have the potential to describe early inflation of the universe. Some modification of the starting action of our models may solve these problems, for instance, considering $\omega$ a function of $\phi$ which gives a Nordtvedt type theory \cite{nord}. We note therefore, in conclusion that the application of the anharmonic oscillator treatment for a varying $\omega$ theory is perhaps the next prescribed step. This may produce novel solutions under the scope of the theory bridging an early inflation, a decelerating radiation and a late time accelerated expansion without violating the astronomical constraints.

\section{Acknowledgment} 
The authors thank Prof. Narayan Banerjee (DPS, IISER Kolkata) and Prof. Sayan Kar (CTS, IIT Kharagpur) for important feedbacks. SC was supported by the National Post-Doctoral Fellowship (file number : PDF/2017/000750) from the Science and Engineering Research Board (SERB), Government of India.

\end{document}